%% file: main.tex
\documentclass[letterpaper,11pt]{article}
\usepackage{latexsym}
\usepackage{graphicx}
\usepackage{mathptmx}
\usepackage[T1]{fontenc}
\usepackage[margin=1in]{geometry}
\usepackage{amsmath}
\usepackage{amsfonts}
\usepackage{amssymb}
\usepackage{amsbsy}
\usepackage{amsthm}
\usepackage{algorithm}
\usepackage{etoolbox}
\usepackage{algorithmic}

\usepackage{hyperref}

\date{}

\input{sections/preamble}

\begin{document}

%
%

\title{Trajectory-oriented optimization of stochastic epidemiological models}

\author{Arindam Fadikar\\ [5pt]
	Decision and Infrastructure Sciences\\
	Argonne National Laboratory\\
\and
Micka\"{e}l Binois\\ [5pt]
Acumes project-team\\
Inria centre at Université Côte d'Azur\\
\and
Nicholson Collier\\ [5pt]
Decision and Infrastructure Sciences\\
Argonne National Laboratory\\
\and
Abby Stevens\\ [5pt]
Decision and Infrastructure Sciences\\
Argonne National Laboratory\\
\and
Kok Ben Toh\\ [5pt]
Department of Preventive Medicine \\
Institute of Global Health\\
Northwestern University\\
\and
Jonathan Ozik\\ [5pt]
Decision and Infrastructure Sciences\\
Argonne National Laboratory\\
}

\newcommand{\mic}[1]{\textcolor{green}{[#1]}}

\maketitle


\input{sections/sec_abstract}

\input{sections/sec_intro}

\input{sections/sec_method}

\input{sections/sec_results}

\input{sections/sec_conclusion}

\input{sections/sec_acknowledgement}

\footnotesize

\bibliographystyle{plain}

\bibliography{allrefs}

\end{document}

%% file: sections/preamble.tex
\usepackage{times}
\usepackage{ifthen}   
\usepackage{color}
\usepackage{subfigure,wrapfig}
\usepackage{multirow}
\usepackage{longtable} 	
\usepackage{datetime}
\usepackage{xspace}
\usepackage{paralist} 
\usepackage{enumitem} 

\usepackage{cleveref} 

\usepackage{titlesec}

\usepackage{mdframed}  
\usepackage{colortbl}

\usepackage{soul}

\makeatletter
\def\namedlabel#1#2{\begingroup
    #2%
    \def\@currentlabel{#2}%
    \phantomsection\label{#1}\endgroup
}
\makeatother

\newcommand{\Sc}{{\mathcal S}}
\newcommand{\T}{{\rm T}} 
\newcommand{\Reals}{\mathbb{R}} 
\newcommand{\R}{\Reals} 
\newcommand{\I}{\mathbb{I}}

\DeclareMathOperator*{\argmin}{arg\,min}

\usepackage{bm} 

\newcommand{\kb}{\bm{k}}
\newcommand{\Kb}{\bm{K}}

\newcommand{\xb}{\bm{x}}
\newcommand{\Xb}{\bm{X}}
\newcommand{\yb}{\bm{y}}
\newcommand{\Yb}{\bm{Y}}

\newcommand{\cX}{\mathcal{X}}

\newcommand{\add}[1]{#1}

\usepackage{tikz}
\usetikzlibrary{positioning}
\usetikzlibrary{calc}
\usetikzlibrary{shapes.geometric}
\usetikzlibrary{topaths}
\usetikzlibrary{external}
\usetikzlibrary{fadings,decorations.pathreplacing} 

\usepackage{pgfplots}
\usetikzlibrary{shapes,decorations.pathmorphing,patterns}
\pgfplotsset{compat=newest}
\usepgfplotslibrary{fillbetween}

\usepackage[flushleft]{threeparttable}

%% file: sections/sec_abstract.tex
\section*{ABSTRACT}

Epidemiological models must be calibrated to ground truth for downstream tasks such as producing forward projections or running what-if scenarios. The meaning of calibration changes in case of a stochastic model since output from such a model is generally described via an ensemble or a distribution. Each member of the ensemble is usually mapped to a random number seed (explicitly or implicitly). With the goal of finding not only the input parameter settings but also the random seeds that are consistent with the ground truth, we propose a class of Gaussian process (GP) surrogates along with an optimization strategy based on Thompson sampling. This \emph{Trajectory Oriented Optimization} (TOO) approach produces actual trajectories close to the empirical observations instead of a set of parameter settings where only the mean simulation behavior matches with the ground truth. 

%% file: sections/sec_intro.tex
\section{INTRODUCTION}
\label{sec:intro}


Stochastic epidemiological models have been shown to be a critical tool to support public health decision-making in times of crisis (see, e.g.,~\cite{Ozik2021}). These models are parameterized by calibrating their output trajectories against observed data, a crucial step for their use in policy-making. Calibration, however, can be computationally infeasible, particularly for complicated simulators and large populations. Rather than requiring simulations to be re-run with all possible parameter combinations, many have proposed using surrogate models for calibration, which replaces the expensive simulation with a simpler model such as a Gaussian process (GP)~\cite{gramacy2020surrogates,carmassi2019bayesian,higdon2004combining,Kennedy2001}.

A typical GP surrogate models the \textit{mean} behavior of a stochastic simulator. Given some initialization, repeated runs of the stochastic simulator will result in an ensemble of random realizations of output trajectories from a distribution (as opposed to a fixed response in the case of deterministic models). In the case of a noisy epidemic model, the stochasticity in the output can safely be represented by a mean and variance (assuming white noise) and a GP surrogate can be produced to predict the mean behavior of this process for each simulation input that, once calibrated, is consistent with the ground truth. In other modeling settings, however, each replicate may represent something more structured than white noise. For example, in the case of an agent-based epidemiological model, each replicate might define a different population mixing, and averaging outputs from all replicates may not correspond to a feasible mixing pattern. 

In such cases, the random seed governing the stochasticity of each replicate can be taken as an additional input to the model, and the calibration procedure is searching not for the input parameters that most closely match the empirical data on average, but rather for the best \textit{trajectory} (parameter-seed pair). This trajectory-oriented optimization (TOO) approach seeks to capture specific mixing patterns in stochastic models, enabling more empirically informed scenario modeling and forward projections. In this paper, we propose a class of trajectory-oriented GP surrogates with an associated optimization strategy. We leverage the common random number framework detailed by, e.g.,~\cite{chen2012effects} and account for non-stationarities with a local formulation of the model~\cite{edwards2021precision}. We develop a flexible Bayesian optimization technique tailored for this specific case using a Thompson sampling approach that can be generalized to multiple targets. \add{The main contribution of this paper is the combined surrogate model and optimization strategy framework for trajectory like response}. We apply our method to an Susceptible-Exposed-Infectious-Recovered \add{(SEIR)} epidemiological model of COVID-19 infections in Chicago, IL, U.S.A., and demonstrate its benefits over standard GP surrogates.

%% file: sections/sec_method.tex
\section{METHOD}
\label{sec:method}

Let us denote the stochastic computer simulation as $f: \R^d \to \R$ that maps a $d$ dimensional input $\xb \in \cX^d \subseteq \R^d$ to a vector output. For stochastic models, $f$ is not one-to-one and the output $f(\xb)$ can be indexed via replicate indices if repeated runs are available at each input. For the purpose of this paper and the trajectory-oriented approach, we augment the $d$-dimensional input $\xb$ with a trajectory identifier $r$ that can also be associated with the random seed used in the simulation. Later, this additional input dimension is used as a categorical input and is exploited in our proposed GP surrogate. In summary, an output from the stochastic computer simulation $f$ would be characterized by the tuple $(\xb, r)$. At this level of specification, $f$ is treated as a deterministic mapping between the tuple $(\xb, r)$ and its output. We also slightly abuse notation in adding the time variable $t$ in $\xb$ to denote time-indexed scalar output from $f$.

A standard GP~\cite{Rasmussen2005,gramacy2020surrogates} prior characterized by a mean function $\mu(\cdot)$ and a covariance function $k(\cdot, \cdot | \Phi)$ on a set of output \add{of size $N$}, $\Yb = (y_1, \ldots, y_N)^\T$ at input locations $\Xb = (\xb_1, \ldots, \xb_N)^\T$ provides a likelihood given by the multivariate Gaussian distribution. The covariance kernel $k: \R^{d+1} \times \R^{d+1} \to \R$ is positive definite and is often parameterized by additional parameter set $\Phi$ that controls various aspects of the input dependency. Inference on this additional unknown parameter is typically done via likelihood maximization or MCMC. A common practice in the computer experiments literature is to replace the actual simulation $f$ by a fitted GP on $N$ experiments for downstream tasks such as optimization, calibration or sensitivity analysis (see e.g.,~\cite{kleijnen2018design,gramacy2020surrogates}). 
When the observations $\Yb$ from a computer experiment are noisy, white noise with fixed variance $\tau^2$ is added to the covariance matrix such that $\Yb \sim MVN(\mu, \Kb_N + \boldsymbol{\Sigma}_N)$, where $\Kb_N$ is the $N\times N$ matrix with $ij$ coordinates $k(\xb_i, \xb_j)$ and $\boldsymbol{\Sigma}_N = \tau^2 \I$. Prediction at a new input $\xb$ conditioning on the observations, i.e., $Y(\xb) | \Yb$ also follows a Gaussian distribution with mean and variance given by:
\begin{align}\label{eq:kriging}
    \mu(\xb) & = \kb(\xb)^\T(\Kb_N + \boldsymbol{\Sigma}_N)^{-1} \Yb, \;\; \text{where} \;\;\kb(\xb) = (k(\xb, \xb_1), \ldots, k(\xb, \xb_N))^\T;\\ \nonumber
    \sigma^2(\xb) & = k(\Xb, \Xb) + \tau^2 -  \kb(\xb)^\T (\Kb_N + \boldsymbol{\Sigma}_N)^{-1} \kb(\xb),
\end{align}
Without loss of generality, we assume $\mu = 0$ for the rest of the description. 


When replicates (repeated experimental runs at the same design point) are available, the above GP formulation remains valid. However, we can achieve computational gain without sacrificing statistical properties by aggregating replicates across inputs and modeling their mean with an updated covariance matrix to account for non-constant variance across inputs. This is often called the stochastic kriging model~\cite{Ankenman2010}, while other options to model input-dependent noise can be found in~\cite{Goldberg1998,Kersting2007,Binois2018,Fadikar2018}.

In this work, however, we are explicitly interested in modeling individual \textit{trajectories} rather than the mean behavior, a desirable task when the random seed encodes some essential structural information into the simulation. During calibration, rather than finding the parameter $\xb$ that best matches observational data in expectation, we search for the optimal parameter-seed pair $(\xb, r)$. 

\subsection{\add{Common Random Number Gaussian Process (CRNGP)}}
We now define a new class of GP surrogates for stochastic computer simulations that models each replicate by exploiting the common random numbers~\cite{chen2012effects,pearce2019bayesian,pearce2022bayesian}. Referring back to the augmented input tuple $(\xb, r)$ for stochastic $f$, we formally define $f$ to be a mapping $\R^d \times \Sc \to \R$, where $\Sc$ is the set of random seeds. In general, $\Sc$ can be arbitrary, but the exact specification is not necessary for its use. Without loss of generality, we assume $\Sc$ to be countable. Now that $f$ is deterministic, we fall back to the standard GP prior with a carefully drafted covariance kernel $k$ to deal with this augmented input. We assume a two-component separable covariance structure here such that, $k((\xb_i, r), (\xb_j, r')) = k(\xb_i, \xb_j) \times k_s(r, r')$. The novelty is in the kernel $k_s$ where we put our assumption(s) on replicate similarity while the kernel $k$ for $\xb$ can be set to a common family of covariance kernels such as Gaussian or Mat\'ern. 

\begin{figure}[ht]
    \centering
    \includegraphics[width = 0.8\textwidth]{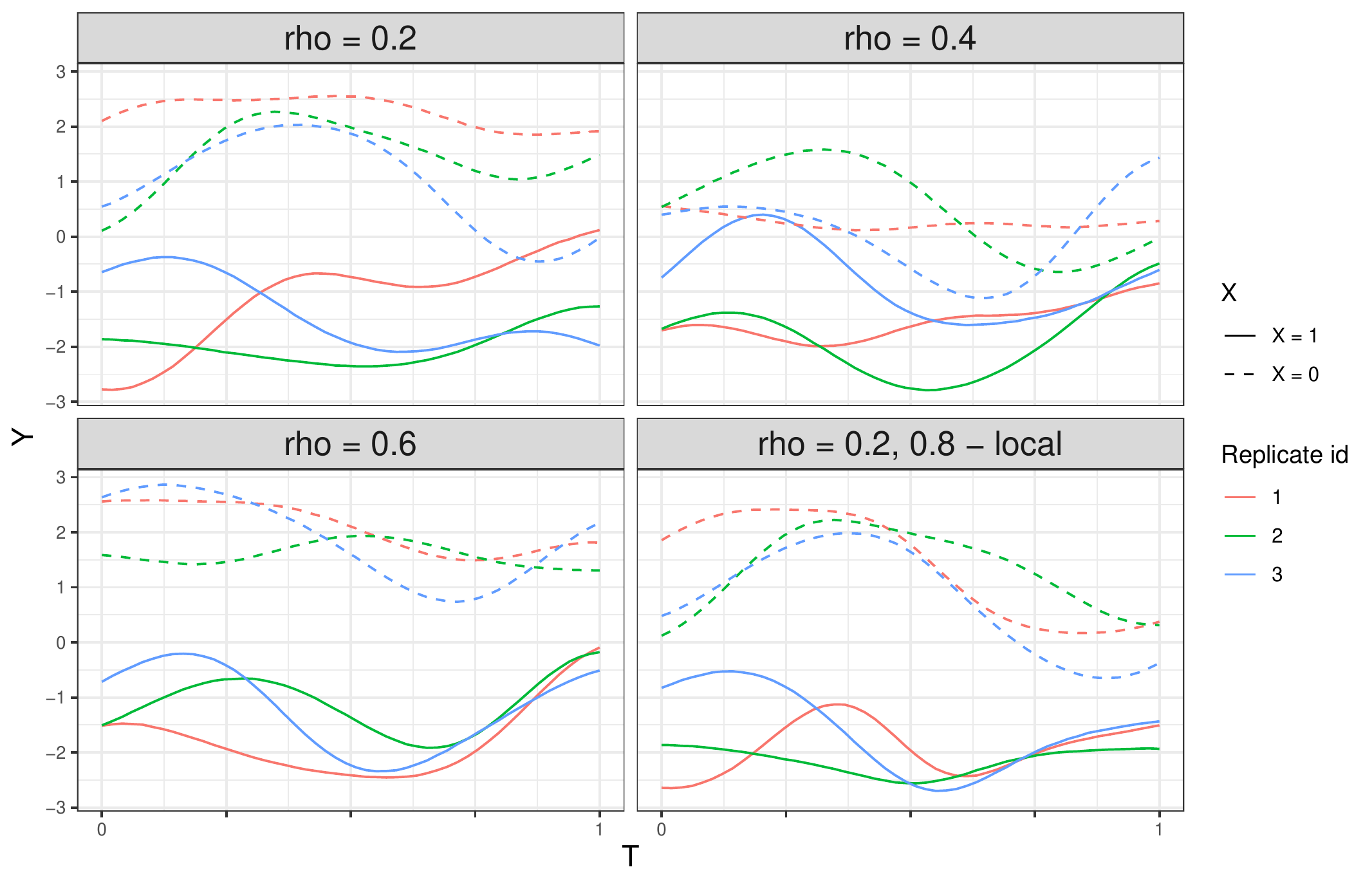}
    \caption{Realizations from CRNGP for different values of $\rho$ (keeping the other GP hyperparameters fixed) except for the bottom-right panel which shows realization from a local CRNGP model. Each line corresponds to a random realization from the CRNGP which is vector output (trajectory) indexed by variable $T$ \add{(scaled to [0, 1]). In practice, $T$ can be thought of as discrete time indices where the simulations are evaluated}. Solid and dashed lines correspond to realizations at $x=1$ and $x=0$ respectively. The local CRNGP consists of a mixture of two CRNGPs with different values of $\rho$s on the partition on $T$-axis $\{ (0, 0.5), (0.5, 1) \}$.}
    \label{fig:crngp_samp}
\end{figure}

Without any prior information on how replicates are related in a given computer simulation, we put a simplified assumption that the marginal covariance between responses at two random seeds $r$ and $r'$ is constant, $k_s(r, r') = \rho$, $0 < \rho < 1$. With this new specification, we write the computer simulation output vector $\Yb = (y_1^{(1)}, \ldots, y_1^{(r_1)}, \ldots, y_n^{(1)}, \ldots, y_n^{(r_n)})$ and the unique input locations $\Xb = (\xb_1, \ldots, \xb_n)$, where, $r_i$ denotes the number of replicates at $\xb_i$ with $\sum_{i=1}^n r_i = N$. Without loss of generality, we also denote the random seeds by the natural numbers since they are treated as categorical in the GP formulation. The elements in covariance matrix $\Kb_N$ is expressed as:
\begin{align}\label{eq:crncov}
    k((\xb_i, r), (\xb_j, r')) = \begin{cases}
        k(\xb_i, \xb_j) & \text{if } r = r'\\
        k(\xb_i, \xb_j) \times \rho & \text{otherwise,} 
    \end{cases}
\end{align}
\add{where $\rho$ can be interpreted as an adjustment to the similarity between responses at $\xb_i$ and $\xb_j$ if different seeds are used at each $\xb$. Moreover, this specification of $\rho$ results in a valid covariance kernel}. A constant nugget $\tau^2$ (typically a small positive number) is added to the diagonal of the matrix $\Kb_N$, i.e.,  $\boldsymbol{\Sigma}_N = \tau^2 \I$. With this formulation, the kriging formula in \Cref{eq:kriging} still holds true with a little modification of the new predicting location $\xb$ which now actually means prediction of a particular replicate $(\xb, r)$. Obtaining the mean behavior (ignoring the seed information) is equivalent to making a prediction at a new seed (i.e., for which no observation has been made)~\cite{pearce2022bayesian}. \Cref{fig:crngp_samp} shows realizations from CRNGP for three different values of $\rho$ where $k$ is set to a Gaussian covariance kernel with length-scale and nugget at 0.01 and 0.001 respectively. The notable observation from this figure is the role of $\rho$ in controlling the similarities among three replicated trajectories. As $\rho$ increases, the trajectories tend to behave similarly at each $\xb$. The bottom-right panel shows a modified version of the proposed CRNGP (hereafter local CRNGP) where the value of $\rho$ depends on the input. We describe this new family of models in detail below.

\begin{figure}[ht]
    \centering
    \includegraphics[width = 0.99\textwidth]{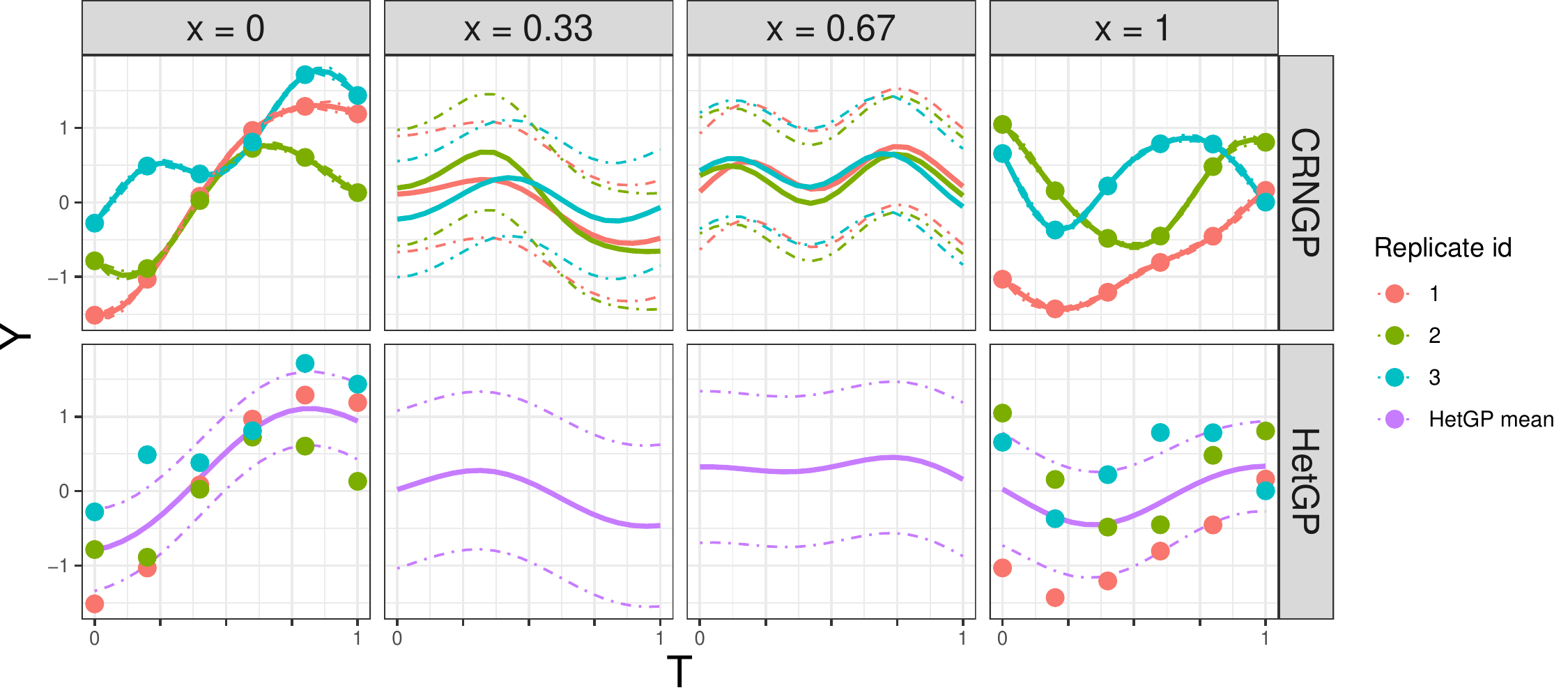}
    \caption{Comparison between fitted CRNGP and Heteroskedastic GP (HetGP) on vector-valued response at 4 different values of input $x$. Each color corresponds to a random seed value except the purple which represents the predicted mean surface from the HetGP model. The top panel shows the prediction from CRNGP which occurs at trajectory level and the bottom panel shows one mean prediction at each $x$ by a heterogeneous GP.}
    \label{fig:crn_vs_hetgp_fit_comp}
\end{figure}

\Cref{fig:crn_vs_hetgp_fit_comp} provides a glimpse of the fundamental difference between CRNGP and other standard GP approaches, we choose heteroskedastic GP~\cite{Binois2018} here for demonstration. When the random seed information is available, CRNGP is perfectly able to interpolate the response surface at the replicate level (for $x=0$ and $x=1$), whereas, a random seed agnostic GP simply tracks the mean behavior of the response at each $(x, T)$. \add{The estimated value of $\rho$ in CRNGP is 0.56}. An optimization algorithm based on just the mean behavior may result in a trajectory that would be otherwise infeasible at any $\xb$.

\textbf{Locally approximated CRNGP}
Note that the modified covariance kernel in CRNGP still maintains the stationarity and hence can be restrictive in real applications. Following common approaches for dealing with non-stationarity via local approximations~(\cite{Gramacy2015a},\cite{chipman2010bart}), we propose a similar recipe to extend the stationary CRNGP to support non-stationary realizations. To relax the assumption that realizations at any $\xb$ has a marginal covariance of $\rho$ that does not depend on the actual value of $\xb$, we start with $L$ partition of the input space $\cX^d = \bigcup_{l = 1}^L \cX^d_{l}$ with $\cX^d_l \cap \cX^d_{l'} = \emptyset$ for any $l \neq l'$. An independent CRNGP is fitted to each subset of the data $(\Xb_l, \Yb_l)$ with its own hyperparameters according to its membership into one of the $L$ bins given the partition. Prediction at a new $\xb^* \in \cX^d_{l^*}$ is followed from the model in $l^*$ partition. While this strategy provides an easy and cheap alternative to a full input-dependent GP hyperparameter estimation, a realization from this locally approximated CRNGP is not smooth and can show unrealistic behaviors at the partition boundaries. Moreover, prediction at a boundary should be influenced by all partitions sharing the boundary than just one partition. For the downstream tasks (such as optimization using Thompson sampling, discussed in the next section), a non-smooth realization on the whole input space may give rise to inaccuracies. 

Instead, we define the predictive distribution at any $\xb^*$ to be a weighted prediction across all $L$ local models where the weights are proportional to the inverse of the distance between $\xb^*$ and $\cX^d_{l}$ for all $l=1, \cdots, L$.
Aggregating local models is also the idea behind product of experts or Bayesian committee machines, which come in many flavors, see, e.g.,~\cite{rulliere2018nested,edwards2021precision} and references therein.
For the purpose of generating realizations from this construction, one can bypass the exact analytic form of the mean and variance of the predictive distribution at $\xb^*$ and directly apply the averaging on realizations from individual models. Let $\Yb^{(l)}(\xb^*, s)$ be the realization from the predictive distribution of $\Yb | (\Xb_l, \Yb_l)$, then the combined realization is given by,
\begin{equation}
    \Yb(\xb^*, s) = \sum_{l=1}^L \alpha(\xb^*, \Xb_l) \Yb^{(l)}(\xb^*, s),
\end{equation}
where the $\alpha$ function depicts the distance between the prediction location and the training data in each partition.

\subsection{Optimization using CRNGP surrogate}


Gaussian processes are the workhorse of Bayesian optimization~\cite{garnett2023bayesian}, also called efficient global optimization~\cite{Jones1998}. The principle is to rely on the prediction and uncertainty quantification capabilities of a GP surrogate to balance between exploitation of pertinent areas where the mean is promising and exploration of unknown regions of the parameter space where the variance is large. This task is handled by the acquisition function or infill criterion, such as the expected improvement~\cite{Mockus1978} for global optimization. We refer to~\cite{garnett2023bayesian} for a more detailed presentation of BO.

In Bayesian optimization, scalar outputs are the most typical use case, with many possible variations. Related to our TOO approach,~\cite{astudillo2021thinking} solve similar trajectory problems by exploiting the composite aspect of the cost function to optimize: $g(\yb(\xb)) =\sum \limits _j (y_j(\xb) - y_j^{obs})^2$ where $y_j^{obs}$ are the actual observations that the simulator trajectory should match. But they solely focus on the mean behavior and not on the actual trajectories. This is also the case of~\cite{pearce2019bayesian,pearce2022bayesian} where they exploit common random numbers with an enriched CRNGP model.


Our goal here is to identify $$\left\{(\xb^{(1)}, r^{(1)}), \dots, (\xb^{(p)}, r^{(p)}) \text{~s.t.~} g(y(\xb^{(i)}, r^{(i)})) \leq g(y(\xb, r^{(j)}) \forall \xb \in \cX \forall 1 \leq i \leq p, 1 \leq j \leq |\Sc| \right \}$$ 
where $p$ is the number of trajectories needed. To do so, we look at the following optimization problem as a proxy: 
\begin{equation}
    \argmin \limits_{\xb \in \cX, r \in \Sc} \sum \limits_j \left(f(\xb, t_j, r) - y_j ^{(obs)} \right)^2
    \label{eq:optprob}
\end{equation} where the set $\Sc$ is countable (in practice, we set $\Sc$ to be a large finite set of random seeds).
Compared to the more usual goal of solving $\argmin \limits_{\xb \in \cX} \mathbb{E}_r \left[ \sum \limits_j \left(f(\xb, t_j, r) - y_j ^{(obs)} \right)^2 \right] = g(\yb(\xb))$, we are not focusing on finding the parameters $\xb$ that perform best on average but on finding good trajectories even if the mean behavior does not correspond to the data. This would allow us to provide a more diverse set of inputs than when focusing solely on the mean. Note that this is also a different approach than focusing on risk-averse (or robust) optimization.

For the trajectory-oriented setup, we propose to use Thompson sampling (TS)~\cite{thompson1933likelihood}, a simple and versatile approach still balancing exploration and exploitation. TS amounts to sampling one realization from the GP posterior distribution $\tilde{\Yb} \sim \mathcal{N} \left(\mu(\tilde{\Xb}, S), \sigma^2(\tilde{\Xb}, S) \right)$ over discrete sets $\tilde{\Xb}$ and $S$, then computing $g(\tilde{\Yb})$ before selecting $ (\xb^*, r^*) \in \argmin \limits_{(\xb \in \tilde{\Xb}, r \in S)} g(\tilde{\Yb})$ as next input to evaluate. Extension to a continuous input space can be performed by using varying random discrete sets or optimizing a continuous approximation of the realization~\cite{Mutny2018}. In practice, as in~\cite{pearce2022bayesian}, the set $S$ only needs to contain seed values that have already been evaluated, plus one additional seed value accounting for the mean behavior over all unobserved seeds. This way, a new seed is evaluated only if it does better than all existing ones while promising seeds can be kept to improve over $\xb$ (\add{line 4 in the~\Cref{alg:CRNBO}}). The general pseudo-code of the approach is illustrated in Algorithm \ref{alg:CRNBO}. The batch extension of this Thompson sampling scheme is performed by sampling several posterior samples and using their respective best solutions as batch candidates~\cite{Kandasamy2018}. \add{The choice of the number of initial simulations--$N_0$ and total simulations--$N_{\max}$ would generally be dictated by the total computational budget available.}

\begin{algorithm}[!ht]
\caption{Pseudo-code for CRNGP based trajectory search}\label{alg:CRNBO}
\begin{algorithmic}[1]
\REQUIRE  $N_0$ (initial budget), $N_{\max}$ (total budget)
\STATE Create initial design of experiments, $(\Xb_n, S_n)$ of size $n = N_0$ and evaluate it $\Yb_n = f(\Xb_n, S_n)$
\STATE Train the GP model on $(\Xb_n,S_n, \Yb_n)$
\WHILE{$n \leq N_{max}$}
    \STATE $S_n \leftarrow \left\{S_n, (\max S_n) + 1 \right\}$ 
    \STATE Sample from the GP posterior: $\tilde{\Yb} \sim \mathcal{N} \left(\mu(\tilde{\Xb}, S_n), \sigma^2(\tilde{\Xb}, S_n) \right)$
    \STATE Choose $(\xb_{n+1}, s_{n+1}) \in \arg \max_{\xb \in \tilde{\Xb}, s \in S_n} g(\tilde{\Yb})$
    \STATE Update the GP model by conditioning on $(\xb_{n+1}, s_{n+1})$.
    \STATE $n \leftarrow n + 1$
\ENDWHILE
\end{algorithmic}
\end{algorithm}

The calibration task may have $p \geq 2$ targets, corresponding to various observables matching outputs of the simulators. In such a case, there is in general no solution perfectly matching all outputs simultaneously, and the solution is defined based on compromise. That is, a solution is said to be Pareto non-dominated if there is no other one that is at least as good on all objectives and strictly better in at least one. The set of non-dominated solutions is called the Pareto set in the input space, its image in the output space being the Pareto front.
In addition to fitting one GP to each output and sampling from their respective posteriors, the sequential selection procedure must be modified as well. For instance, the hypervolume contribution, e.g.,~\cite{beume2007sms} of each element of the GP posterior, i.e., $\left(g(\tilde{\Yb}_1^{(i)} \right), \dots, g(\tilde{\Yb}_p^{(i)}))$, is the volume added to the current Pareto formed by the  $(g(y_1^{(i)}), \dots, g(y_p^{(i)}))_{1 \leq i \leq n}$. 


\begin{figure}[!ht]
    \centering
    \includegraphics[width = 0.99\textwidth]{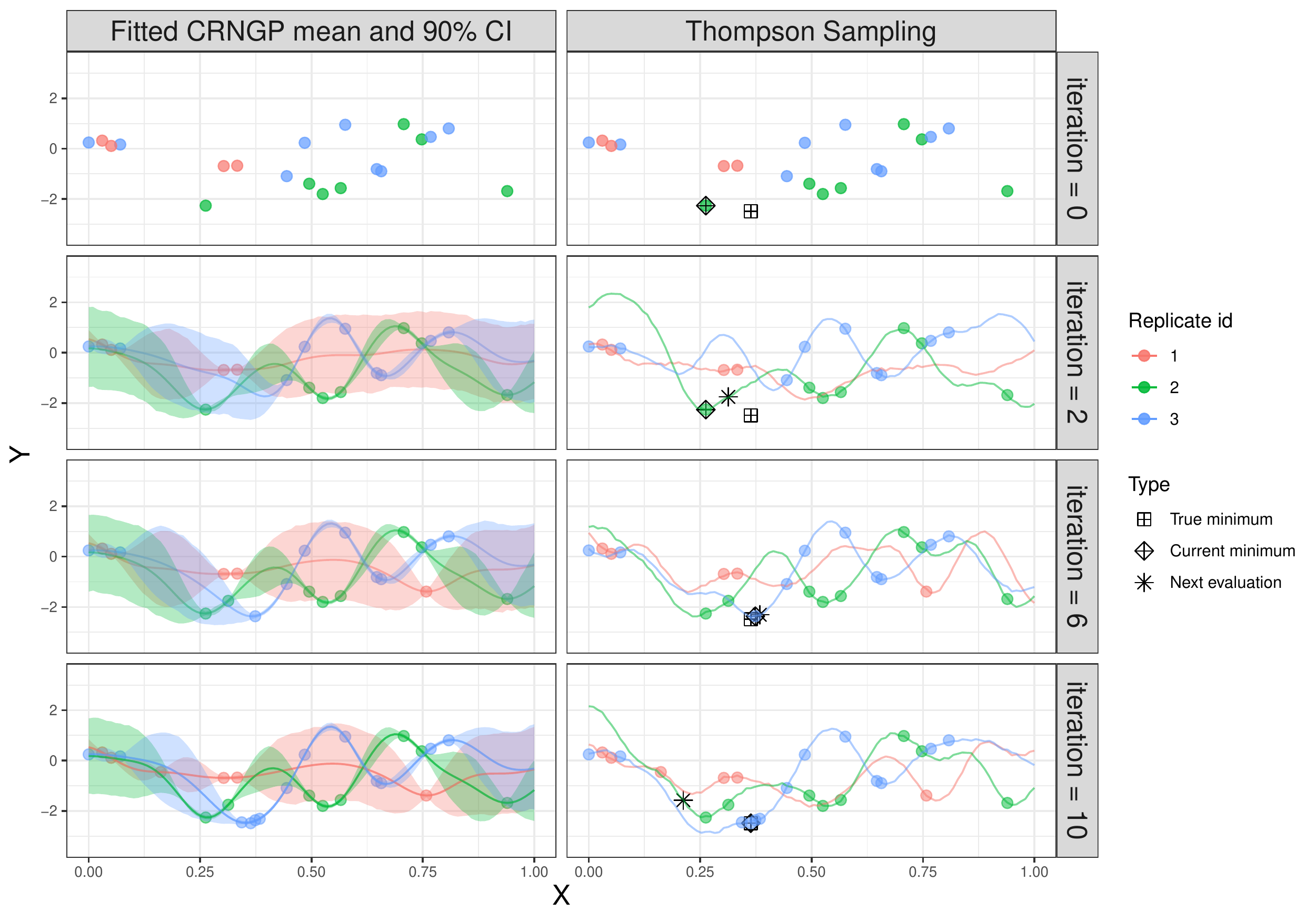}
    \caption{Thompson sampling (TS) based single objective optimization using CRNGP surrogate. Figures in the left show the predicted mean surface by CRNGP conditioning on the observations (represented by dots) and the figures in the right show random realizations from the fitted CRNGP based on which the new evaluations are obtained in the TS algorithm. Replicates are represented by different colors. Rows correspond to different iterations in the TS. In each iteration a new $x$ for next evaluation is set at the arg min of a realization (right column). The global minimum is found in 11 iterations.}
    \label{fig:crn_TS}
\end{figure}

%% file: sections/sec_results.tex
\section{Empirical evaluation}
\label{sec:result}

\subsection{Stochastic Epidemic Model}

We used a published stochastic SEIR compartmental model \cite{runge2022modeling} to simulate SARS-CoV-2 transmission and COVID-19 disease states. The model included multiple symptom statuses (asymptomatic, presymptomatic, mild, and severe), and severe disease outcomes (requiring hospitalization, critical illness requiring intensive care unit ICU admission, and deaths). 

\begin{figure}[htpb]
\centering
\includegraphics[scale=1]{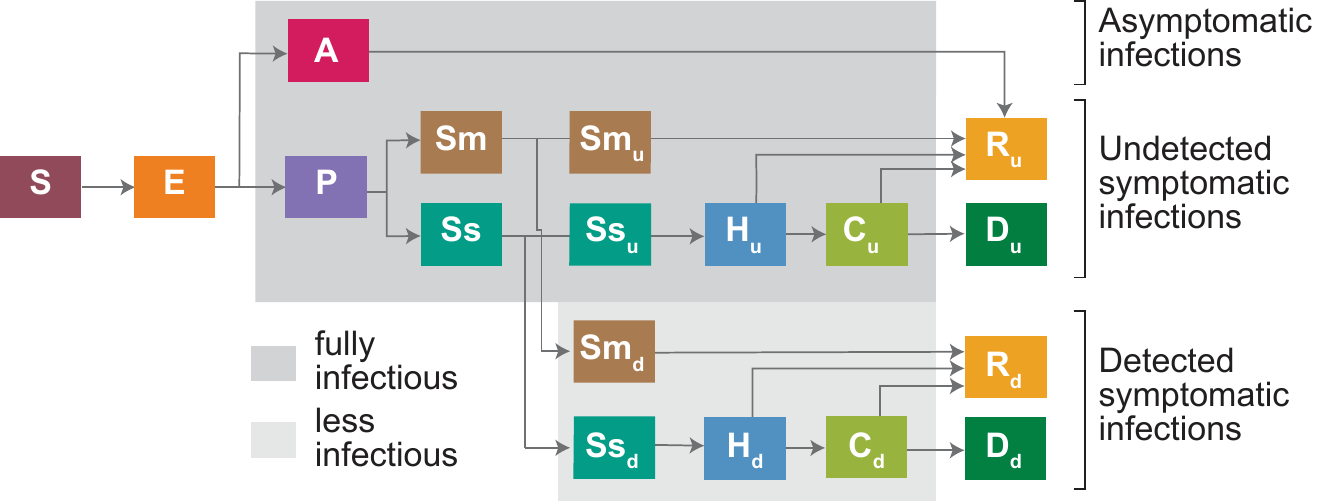}
\caption{{\bf SEIR model structure to simulate SARS-CoV-2 transmission.}
 S=Susceptible, E=Exposed, A=Asymptomatic, P=Presymptomatic, Sm=Mild symptomatic, Ss=Severe symptomatic, H=Requires hospitalization, C=Critically ill (ICU), D=Death, R=Recovered. Subscripts u and d denote if the infection is undetected or detected. Asymptomatic and presymptomatic infections may also be detected, which would result in less infectiousness.}
\label{fig:seir}
\end{figure}

At the start of the epidemic, everyone is susceptible (S). The susceptible population is exposed and infected by the virus at a rate controlled by the transmission rate parameter, the number of infectious and susceptible people. After a latent period of a few days, exposed individuals (E) become infectious without symptoms: they are either completely asymptomatic (As) or presymptomatic (P). At the end of the incubation period, presymptomatic people proceed to develop either mild (Sm) or severe (Ss) symptoms. Asymptomatic or people with mild symptoms recovered (R) eventually, while those with severe symptoms would require hospital care (H) after a few days in the Ss compartment. Whether or not they eventually sought care, people who need hospital care either recover without complications or develop critical illness (C) requiring ICU care. The latter could eventually recover from the illness by first moving back to the post-ICU hospitalization state (Hp) or die (D) from the illness.  All exposed people are undetected.  When a person enters the asymptomatic, presymptomatic, symptomatic mild, and symptomatic severe state, a fraction of them would be detected after a few days, which leads to isolation and reduced infectiousness. A schematic representation of the model is provided in \Cref{fig:seir}.

Model parameters such as incubation period, disease symptomaticity and severity, and hospital and ICU lengths of stay, are informed by literature and observed data in Chicago.  The transmission rate, probability of detection, and the disease recovery rate are varied over time to better fit the daily ICU census, hospital census, and reported deaths in Chicago. \add{A detailed description of the model and the parameters along with the source code can be found at \href{https://github.com/numalariamodeling/covid-chicago}{this github repository} \cite{covidmodel}}.

\subsection{Single and Bi-objective Optimization}

To illustrate our method, we use a three parameter version of the stochastic compartmental model described above. The three input parameters are the initial disease transmissibility rate and the rate of exposed individuals becoming infectious for both asymptomatic and symptomatic cases. Other disease parameters in the model are kept fixed at values informed by literature and the observed data in Chicago. Each simulation is run for the first 100 days of the disease propagation and output is obtained at 5 discretized time values, essentially resulting in a vector-valued output of size 5. For the output of interest, we choose cumulative hospital admissions in the single objective case along with cumulative deaths in the bi-objective case. A trajectory corresponding to a random (input, random seed) pair is treated as the ground truth and we set our goal to discover all trajectories that are consistent with this ground truth. In each case we set our initial simulation budget $N_0 = 50$, which is uniformly distributed on the three dimensional input space according to a Latin hypercube design \cite{mckay2000comparison} and maximum budget of $N_{\text{max}}=200$.

\begin{figure}[!ht]
    \centering
    \includegraphics[width = 0.7\textwidth]{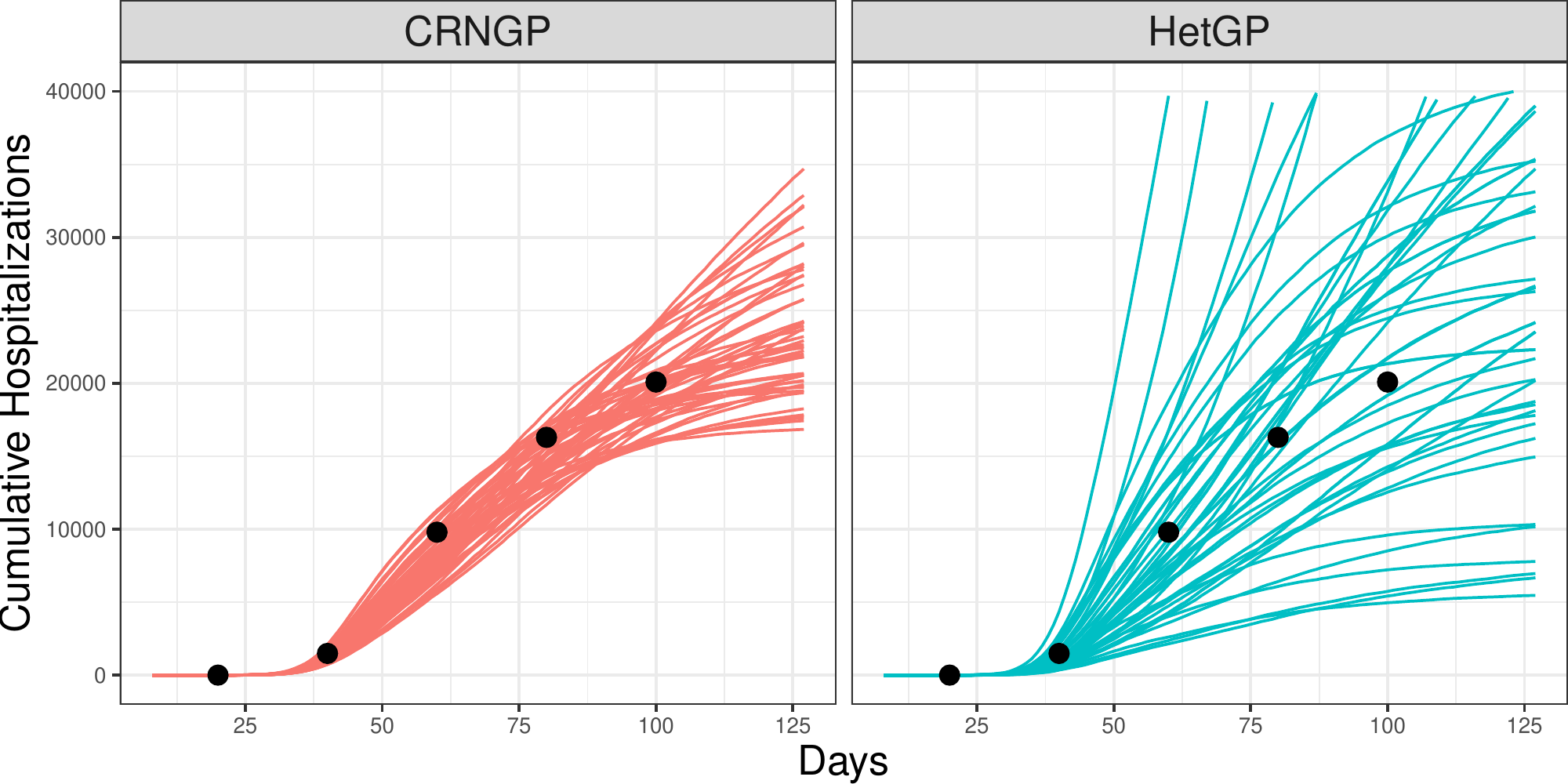}
    \caption{Best 50 trajectories of cumulative hospitalizations that are consistent with the observations denoted by black dots when using CRNGP and HetGP surrogates in Algorithm \ref{alg:CRNBO}.}
    \label{fig:crn_vs_hetgp_1d}
\end{figure}

\begin{figure}[!ht]
    \centering
    \includegraphics[width = 0.7\textwidth]{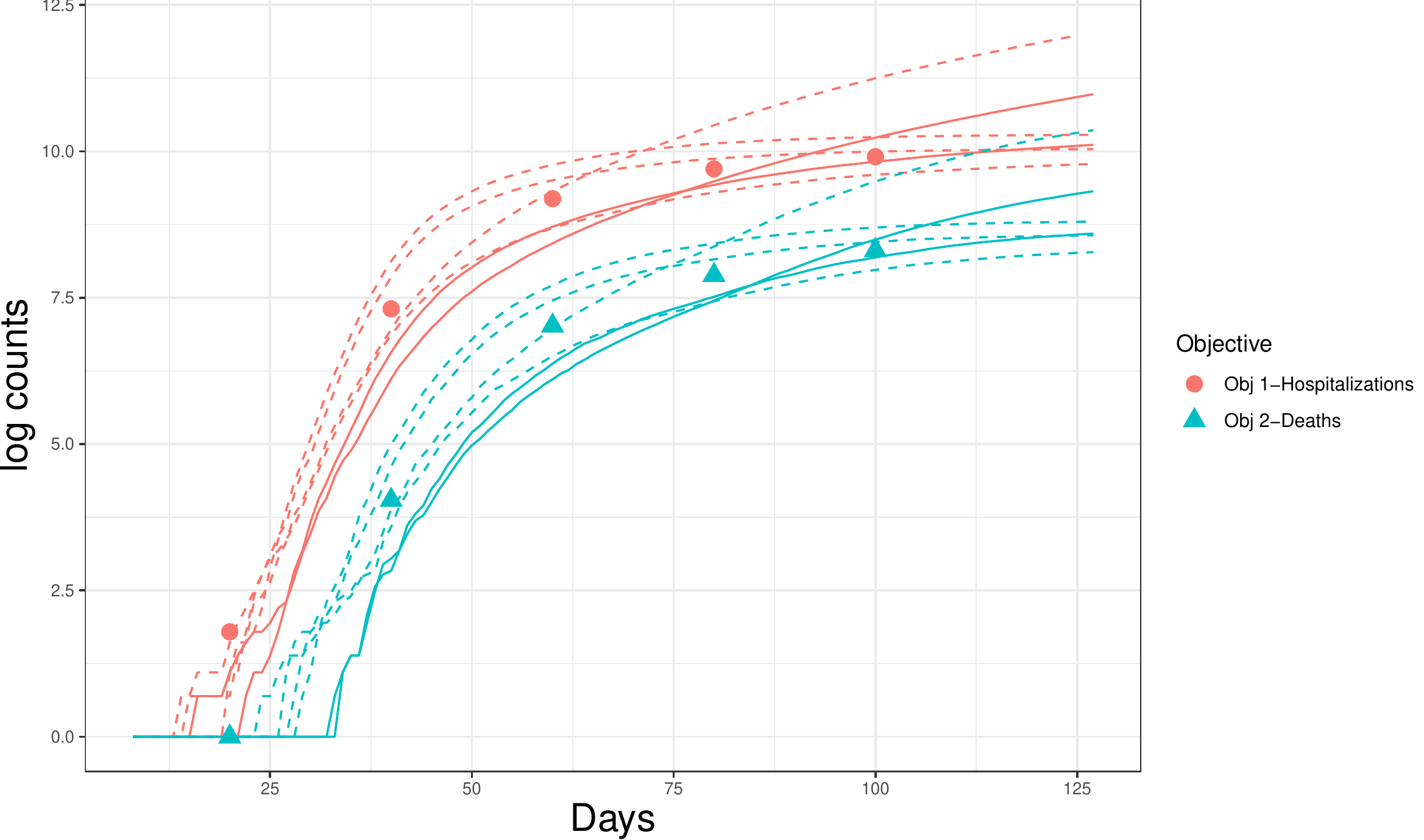}
    \caption{Best trajectories with respect to two objectives--cumulative hospitalizations and deaths simultaneously. Solid and broken lines represent CRNGP and HetGP surrogates respectively.}
    \label{fig:crn_vs_hetgp_pf}
\end{figure}

We also use the local version of the CRNGP surrogate, based on a fixed partition on the time component of the input. Essentially our CRNGP surrogate for vector valued output operates on five dimensional input space -- three disease model parameters, discretized time indices, and the random seed. We choose only the time dimension for constituting a partition on which local surrogates would be more appropriate. This is further evident from \Cref{fig:crn_vs_hetgp_1d} and \Cref{fig:crn_vs_hetgp_pf} as to how trajectories become more diverse with time. The other aspect of a useful partition for local surrogates has to do with the number of actual simulations in each partition. A surrogate trained on only a few simulations in each partition in a high input dimension setting may defeat the goal of creating an accurate surrogate.

\Cref{fig:crn_vs_hetgp_1d} shows the 50 best trajectories that match with the ground truth (denoted by solid black dots) after the total simulation budget is exhausted. Searching for optimal input parameter values using HetGP surrogate in TS results in settings where the mean behavior of the stochastic simulation is expected to match the ground truth, whereas, the TOO approach facilitates finding not only the optimum input parameter values but also the replicates, i.e., an actual trajectory that matches with the ground truth. When calibrating to two objectives -- hospitalizations and deaths, \Cref{fig:crn_vs_hetgp_pf} shows the best trajectories with respect to both quantity of interest.


%% file: sections/sec_conclusion.tex
\section{CONCLUSION}

In the expensive simulation context, see, e.g.,~\cite{baker2022analyzing}, where simulation time is not negligible, the use of surrogate models such as Gaussian processes is required to efficiently obtain a solution of an inverse problem within a limited computational budget. For a simulation that produces a stochastic response, often characterized by a random seed, we have proposed the CRNGP surrogate that can map each individual simulation response to a unique (input, random seed) pair. Furthermore, we have devised an optimization scheme that when paired with CRNGP can facilitate finding actual simulation outputs that are consistent with observations. The proposed method has been developed for handling vector-valued output, hence we use the term \emph{trajectory-oriented optimization}.

In our specific epidemic application, the end goal of calibrating the simulation model is to collect simulated trajectories of disease outcomes that are consistent with empirical data trajectories. For a stochastic epidemic model (such as an ABM), each output of the simulation model may correspond to a specific instantiation of population mixing, which is often not controlled by the input parameters, but by the random seeds. Thus, in order to capture mixing patterns consistent with empirical data, one needs to find both optimal parameter settings and their corresponding random seeds. Specific mixing patterns are important to try to capture since they govern the pathways through which an epidemic will transmit, affecting how public health interventions may mitigate that transmission.
Separately, in order to implement data assimilation for an ABM without rerunning the model from the very beginning of an epidemic, the state of the model, including the complete state of the agents in the model, needs to be captured at the end of each assimilating time period, so that it can be restarted from that time. The TOO approach explicitly seeks to find the best trajectories to keep in this case.

We have shown that using a CRNGP, we are not only able to find good trajectories, but we can do so within a smaller simulation budget. This fast time-to-solution aspect is particularly critical for producing actionable insights for decision makers and has been highlighted in~\cite{Binois2021portfolio}. Further improvement to time-to-solution is possible by considering the simulation time as an additional variable to control (see~\cite{Snoek2012}, \cite{Picheny2013b}) and by developing open science platforms~\cite{collier_developing_2023} to enable these analyses. The local CRNGP surrogate can also be extended to handle input-dependent replicate similarity by considering a more complex prior model for $\rho$.





%% file: sections/sec_acknowledgement.tex
\section{ACKNOWLEDGEMENTS}

This material is based upon work supported by the National Science Foundation under Grant No. 2200234, the National Institutes of Health under grants R01AI136056 and R01DA055502, the U.S. Department of Energy, Office of Science, under contract number DE-AC02-06CH11357, and the DOE Office of Science through the Bio-preparedness Research Virtual Environment (BRaVE) initiative.
This research was completed with resources provided by the Laboratory Computing Resource Center at Argonne National Laboratory (Bebop cluster).